\documentclass[a4paper]{article}

\usepackage[numbers,sort&compress]{natbib} 
\usepackage[utf8]{inputenc}

\usepackage{amsmath}
\usepackage{amsfonts}
\usepackage{amssymb}
\usepackage{mathtools}   
\usepackage{dsfont}

\usepackage{stmaryrd}

\usepackage{comment}
\usepackage{enumerate}

\usepackage{myprobdescr}
\usepackage[backgroundcolor=gray!10,linecolor=gray,bordercolor=black]{todonotes}

\usepackage{authblk}

\usepackage{amsthm}
{\bfseries}{\normalfont}
{\bfseries}{\normalfont}
{\bfseries}{\normalfont}
{\bfseries}{\normalfont}
{\bfseries}{\normalfont}
{\bfseries}{\normalfont}
{\bfseries}{\normalfont}

\usepackage{tikz}
\usetikzlibrary{arrows,decorations.pathmorphing,decorations.pathreplacing,backgrounds,positioning,fit}
\usetikzlibrary{shapes,calc}

\usepackage{microtype,ellipsis}

\usepackage[pagebackref,pdfdisplaydoctitle,menucolor=orange!40!black,filecolor=magenta!40!black,urlcolor=blue!40!black,linkcolor=red!40!black,citecolor=green!40!black,colorlinks]{hyperref}

\def\papertitle{Parameterized Algorithmics \\ for Graph Modification Problems: \\ On Interactions with Heuristics}

\hypersetup{pdftitle={Parameterized Algorithmics for Graph Modification Problems: On Interactions with Heuristics}, pdfauthor={Christian Komusiewicz, André Nichterlein, and Rolf Niedermeier}}

\usepackage{algorithm}
\usepackage{units}

\makeatletter
\def\NAT@spacechar{~}
\makeatother

\usepackage[sort&compress]{cleveref}

\usepackage{etoolbox}

\newcommand{\appendixproof}[2]{
}

\newcommand{\FPT}{\textsf{FPT}\xspace}

\crefname{rrule}{Rule}{Rules}
\crefname{observation}{Observation}{Observations}
\crefname{line}{Line}{Lines}
\crefname{equation}{Equality}{Equalities}
\Crefname{equation}{Inequality}{Inequalities}
\crefname{section}{Section}{Sections}
\crefname{subsection}{Subsection}{Subsections}
\crefname{enumi}{Property}{Properties}

\title{\papertitle}

\newcommand{\N}{\mathds{N}}

\newcommand{\probDefEnv}[1]{%
  \begin{center}%
    \begin{minipage}{0.95\linewidth}%
		#1
    \end{minipage}%
  \end{center}%
}

\author[]{Christian Komusiewicz}
\author[]{André Nichterlein}
\author[]{Rolf Niedermeier}

\affil[]{
  Institut für Softwaretechnik und Theoretische Informatik, TU Berlin, Germany\\ 
  \texttt{\{christian.komusiewicz,andre.nichterlein,rolf.niedermeier\}@tu-berlin.de}
}

\date{}

\newcommand{\kDegAnon}{\textsc{Degree Anonymity}\xspace}

\newcommand{\inputSolution}[8]{

	\begin{tikzpicture}[draw=black!75,>=stealth']
		\node[anchor=north west] (input) at (#3 + 0.05,#1 - 0.05) {\textbf{Input:}};
		\draw[rounded corners] (input.north west)+(-0.05,+0.05) rectangle (#4,#2);
		\tikzstyle{vertex}=[circle,draw=black!80,minimum size=12pt,inner sep=0pt]
		#7
		
		\pgfmathsetmacro{\value}{#4 - #5 + 0.5};
		\begin{scope}[xshift=\value cm]
			\node[anchor=north west] (solution) at (#5 + 0.05,#1 - 0.05) {\textbf{Solution:}};
			\draw[rounded corners] (solution.north west)+(-0.05,+0.05) rectangle (#6,#2);
			#8
		\end{scope}
	\end{tikzpicture}
}

\newcommand{\illustratedProbDefEnv}[2]{%
  \begin{center}%
  \hbadness=10000
    \begin{minipage}{0.95\linewidth}%
		#1
    \end{minipage}
    \smallskip
    #2
  \end{center}%
}

\newcommand{\tworows}[2]{\begin{tabular}{c}{#1}\\{#2}\end{tabular}}

\def\n{5}
\def\kValue{4}
\def\scale{1}

\begin{document}

\maketitle

\begin{abstract}
In graph modification problems, one is given
a graph~$G$ and the goal is to apply a minimum number 
of modification operations (such as edge deletions) to~$G$ such 
that the resulting graph fulfills a certain property. 
For example, the \textsc{Cluster Deletion} problem asks to delete as few
edges as possible such that the resulting graph is a disjoint union of cliques.
Graph modification problems appear in numerous applications, 
including the analysis of biological and social networks. 
Typically, graph modification problems are NP-hard, making them 
natural candidates for parameterized complexity studies.
We discuss several fruitful 
interactions between the development of fixed-parameter algorithms
and the design of heuristics for graph modification problems,
featuring quite different aspects 
of mutual benefits.
\end{abstract}


\section{Introduction}\label{sec:intro}
Graph modification problems lie in the intersection of algorithmics, graph
theory, and network analysis.\footnote{Also refer to the 2014~Dagstuhl 
Seminar~14071 on ``Graph Modification Problems'' organized by Hans L.~Bodlaender,
Pinar Heggernes, and Daniel Lokshtanov~\cite{BHL14}. 
Liu et al.~\cite{LWG14} survey kernelization algorithms for graph modification 
problems.} Formally, a graph modification problem is given as follows.
\probDefEnv{\defDecprob{Graph Modification}
	{A graph~$G=(V,E)$, a graph property~$\Pi$, and an integer~$k \in \N$.}
	{Can~$G$ be transformed with at most~$k$~modification operations into a graph satisfying~$\Pi$?}}
Herein, graph modification operations include edge deletions, insertions, and 
contractions, and vertex deletions. Classic examples for~$\Pi$
are ``being edgeless'' (this is known as \textsc{Vertex Cover}
when the allowed modification operation is vertex deletion)
and ``being a disjoint union of cliques'' (this is known as
\textsc{Cluster Editing} when the allowed modification operations
are edge deletion and insertion).
 
We will deal with simple and natural
graph modification problems that are motivated by real-world
applications. In these applications, the common way of solving these
problems is via heuristics. 

We present four main themes on how the interaction between parameterized 
algorithmics and heuristics can take place, each time illustrated
by some ``key'' graph modification problems.

In Section~\ref{sec:hcd}, we consider a graph-based clustering problem
that has been defined only implicitly by means of a greedy heuristic~\cite{HS00}.
We describe how a natural NP-hard parameterized problem (referred
to as \textsc{Highly Connected Deletion}) can be derived 
from this, and how this leads to further insight into the corresponding 
clustering approach~\cite{HKLN14}.

In Section~\ref{sec:deganon}, starting with a practically successful
heuristic for anonymizing social networks~\cite{LT08} (the corresponding NP-hard
problem is known as~\textsc{Degree Anonymity}), we describe how a
closer inspection yields that either the corresponding approach provides
optimal solutions in polynomial time or one can derive a polynomial-size 
problem kernel with respect to the parameter maximum vertex degree of 
the underlying graph~\cite{HNNS15}. Moreover, we briefly indicate 
how this led---in a feedback loop, so to speak---to improvements also for the heuristic approach~\cite{HHN14}.

In Section~\ref{sec:fast}, we study parameterized local
search---the parameter is the degree of locality~\cite{FFLRSV12}.
Local search is a key technique in combinatorial optimization and the
design of ``improvement heuristics''.  We address both limitations and
prospects of this approach. We discuss, among others, the NP-hard
example problems \textsc{Vertex Cover} and \textsc{Feedback Arc Set in
  Tournaments}.

In Section~\ref{sec:pbo}, we finally discuss how 
one may speed up parameterized algorithms by a clever use of heuristics.
In particular, we discuss parameterization above lower bounds derived 
from linear programming relaxations~\cite{LNRRS14} (here the key example 
is the NP-hard \textsc{Vertex Cover} problem), and the idea of programming by 
optimization~\cite{HH15,Hoo12} (here the key example is the NP-hard
\textsc{Cluster Editing} problem).
We draw some final conclusions in Section~\ref{sec:concl}.

\paragraph{Preliminaries.}
We assume familiarity with fundamental concepts of 
graph theory, algorithms, and complexity. 

A \emph{parameterized problem} is a set of instances of the form $({\cal I}, k)$, where  ${\cal I} \in \Sigma^*$ for a finite alphabet $\Sigma$, and $k \in \N$ is the \emph{parameter}.
A parameterized problem~$Q$ is \emph{fixed-parameter tractable}, shortly \FPT, if there exists an algorithm that on input  $({\cal I}, k)$ decides whether $({\cal I}, k)$ is a yes-instance of~$Q$ in time $f(k)\cdot|{\cal I}|^{O(1)}$,  where $f$ is a computable function independent of $|{\cal I}|$.
A parameterized problem $Q$ is \emph{kernelizable} if there exists a polynomial-time algorithm that maps an instance $({\cal I},k)$ of $Q$ to an instance $({\cal I}',k')$ of $Q$ such that
$|{\cal I}'| \leq \lambda(k)$ for some computable function
     $\lambda$,
$k' \leq \lambda(k)$, and
$({\cal I},k)$ is a yes-instance of $Q$ if and only if
     $({\cal I}',k')$ is a yes-instance of $Q$.
The instance $({\cal I}',k')$ is called a \emph{kernel} of $({\cal I}, k)$.

A problem that is W[1]-hard does not admit a fixed-parameter algorithm, unless the widely believed conjecture FPT${}\neq{}$W[1] fails.

\section{From Heuristics to Parameterized Problems}\label{sec:hcd}
In the following, we illustrate how the consideration of heuristic
algorithms may lead to the definition of new interesting graph
modification problems. 
Our example concerns the interplay between two standard approaches for graph-based data clustering.
%

One approach is to formalize desired properties of clusters and then
to find a clustering of the graph such that the output clusters fulfill
these properties. This clustering can be obtained by modifying the
input graph for example by deleting edges so that all remaining edges
are only inside clusters. Starting with \textsc{Cluster
  Editing}~\cite{GGHN05}, there are by now numerous parameterized
algorithmics studies on graph modification problems related to
clustering, varying on the cluster graph
definition~\cite{BFHMPR10,DM14,FGKNU11,GKNU10,LZZ12}, the modification
operation~\cite{HKMN10}, or both~\cite{BMN12}.
Most of the examples of variants of \textsc{Cluster Editing} evolved
primarily from a graph-theoretic interest.

Another approach is to define the clustering algorithmically, that is,
to describe an algorithm that outputs a clustering and to analyze the
properties of the clusters that are produced by the algorithm. In this
section, we discuss how the consideration of a popular and natural
clustering algorithm due to~\citet{HS00} leads to the definition of the
graph modification problem \textsc{Highly Connected Deletion}. This is
our key example for how to obtain practically motivated parameterized
graph modification problems by a closer inspection of known
heuristics. The study of this new problem then may yield new
challenges for parameterized algorithmics and, furthermore, provide a
better understanding of the strengths and weaknesses of the original
heuristic algorithms.  We will first discuss the original algorithm
and then how we obtain the definition of \textsc{Highly Connected
  Deletion} from this algorithm.

\Citet{HS00} posed the following connectivity
demands on each cluster: the \emph{edge connectivity} $\lambda(G)$
of a graph~$G$ is the minimum number of edges whose deletion results
in a disconnected graph, and a graph~$G$ with $n$ vertices is called
\emph{highly connected} if $\lambda(G) > n/2$.\footnote{An equivalent
characterization is that a graph is highly connected 
if each vertex has degree greater than~$n/2$~\cite{Cha66}.} 
The algorithm by \citet{HS00} partitions the vertex set of the given graph
such that each partition set is highly connected by 
iteratively deleting the edges of a minimum cut in a connected
component that is not yet highly connected. The output clusters of the
algorithm are the connected components of the remaining graph which
are then highly connected. The definition of being highly connected
ensures several useful cluster properties, for example that at least
half of the possible edges are present within each cluster and that
each cluster has diameter at most two~\cite{HS00}.

While Hartuv and Shamir's algorithm guarantees to output a
partitioning into \emph{highly connected} subgraphs, it iteratively
uses a greedy step to find small edge sets to
delete. 
As a consequence, it is not ensured that the
partitioning comes along with a \emph{minimum number of edge
  deletions} making the resulting graphs consist of highly connected
components.  This naturally leads to the edge deletion problem
\textsc{Highly Connected Deletion} where the goal is to
minimize the number of edge deletions; this optimization goal is addressed only implicitly by Hartuv and Shamir's algorithm.
\illustratedProbDefEnv{\defDecprob{\textsc{Highly Connected Deletion}}
	{An undirected graph~$G = (V, E)$ and an integer $k\in\N$.}
	{ Is there an edge set~$E' \subseteq E$ of size at most~$k$
    such that in $G' = (V, E \setminus E')$ all connected components
    are highly connected?} }
{
	\inputSolution{\scale + 0.75}{-\scale - 0.25}{-3.75}{1.75}{-3.75}{1.75}{
		\node[anchor=north west] at (0,\scale + 0.7) {{$k=3$}};
		\tikzstyle{knoten}=[circle,draw,fill=black!20,minimum size=5pt,inner sep=2pt]	

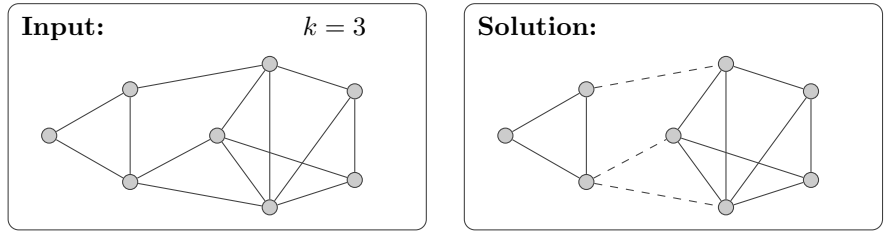
                \foreach \i in {1,...,3} {
                  \node[knoten] (T-\i) at ({0.7\scale * cos(360 * \i / 3 - 3*(180 / 3))-2.5},{0.7\scale * sin(360 * \i / 3 - 3*(180 / 3))}) {};
                }
                	\path (T-1) edge[-] (T-2);
                	\path (T-2) edge[-] (T-3);
                	\path (T-3) edge[-] (T-1);

		\foreach \i in {1,...,\n} {
                  \node[knoten] (K-\i) at ({\scale * cos(360 * \i / \n - 3*(180 / \n))},{\scale * sin(360 * \i / \n - 3*(180 / \n))}) {};
                }
                \foreach \i in {2,...,\n}{
                  \pgfmathtruncatemacro{\value}{\i - 1};
                  \path (K-\i) edge[-] (K-\value);
		}
		\foreach \i / \j in {1/4,1/5,2/5,3/5}{
                  \path (K-\i) edge[-] (K-\j);
		}

                \path (T-1) edge[-] (K-5);
                \path (T-2) edge[-] (K-3);
                \path (T-1) edge[-] (K-4);
	}{


                \foreach \i in {1,...,3} {
                  \node[knoten] (T-\i) at ({0.7\scale * cos(360 * \i / 3 - 3*(180 / 3))-2.5},{0.7\scale * sin(360 * \i / 3 - 3*(180 / 3))}) {};
                }
                	\path (T-1) edge[-] (T-2);
                	\path (T-2) edge[-] (T-3);
                	\path (T-3) edge[-] (T-1);

		\foreach \i in {1,...,\n} {
                  \node[knoten] (K-\i) at ({\scale * cos(360 * \i / \n - 3*(180 / \n))},{\scale * sin(360 * \i / \n - 3*(180 / \n))}) {};
                }
                \foreach \i in {2,...,\n}{
                  \pgfmathtruncatemacro{\value}{\i - 1};
                  \path (K-\i) edge[-] (K-\value);
		}
		\foreach \i / \j in {1/4,1/5,2/5,3/5}{
                  \path (K-\i) edge[-] (K-\j);
		}
                \path (T-1) edge[dashed] (K-5);
                \path (T-2) edge[dashed] (K-3);
                \path (T-1) edge[dashed] (K-4);
	}
}


Interestingly, in the worst case 
the algorithm by \citet{HS00} does not give a good approximation for the
optimization version of \textsc{Highly Connected Deletion}. 
Consider two cliques with
vertex sets $u_1, \dots, u_n$ and $v_1, \dots, v_n$, respectively, and
the additional edges $\{u_i, v_i\}$ for $2 \leq i \leq n$. Then these
additional edges form a solution set of size $n - 1$; however,
Hartuv and Shamir's algorithm will (with unlucky choices of minimum cuts)
transform one of the two cliques into
an independent set by repeatedly cutting off one vertex, thereby
deleting $n(n+1)/2-1$ edges. 
%
 
The following theoretical results are known
for \textsc{Highly Connected Deletion}~\cite{HKLN14}.
It is NP-hard even on
4-regular graphs and, provided the Exponential Time Hypothesis
(ETH)~\cite{IPZ01} is correct, cannot be solved in subexponential
time. On the positive side, there is 
a kernelization that can in polynomial time reduce
an instance to one containing at most~$10 \cdot k^{1.5}$ vertices, and
an FPT~algorithm that solves \textsc{Highly Connected
  Deletion} in $O(3^{4k} \cdot k^2 + n^{O(1)})$ time.

As to the relevance of parameterized algorithmics for
\textsc{Highly Connected Deletion}, one has to note that the 
mentioned FPT~algorithm is impractical. 
In terms of exact solutions, an integer linear programming formulation combined with data reduction rules (partially coming from the kernelization results), however, performs reasonably well~\cite{HKLN14}.
Even when relaxing the goal to find exact solutions
for \textsc{Highly Connected Deletion}, data reduction turned out to 
be beneficial in combination with heuristics 
(improving running time and solution quality)~\cite{HKLN14}.
In a nutshell, the most practical contribution of 
parameterized algorithmics in this example is the development of 
efficient and effective data reduction rules, also helping to improve inexact
solutions based on heuristics. A further benefit of considering a formally defined edge modification problem 
\textsc{Highly Connected Deletion} is that the objective is now independent of a heuristic method used to find it.
Thus, it becomes possible to evaluate the biological quality of the objective~\cite{HKLN14}.

As to potential for future research with respect to \textsc{Highly Connected Deletion}, so far other modification
operations combined with the used cluster graph model 
are unexplored.  Improvements on
the known kernelization for \textsc{Highly Connected Deletion} may
have direct practical impact. Moreover, a first step to make the
FPT~algorithm more practical could be to devise a faster
FPT~algorithm that relies only on branching (the current
algorithm uses dynamic programming in a subroutine).  Finally, besides
striving for improvements with respect to the standard parameter
``number of edge deletions'', the investigation of other
parameterizations may be interesting as well.

{F}rom a more general perspective, however, it remains to ``remodel''
further heuristic algorithms into natural parameterized problems.

\section{Interpreting Heuristics with FPT Methods}\label{sec:deganon}
While in the previous section we derived a natural parameterized 
problem (\textsc{Highly Connected Deletion}) from a simple and effective
greedy heuristic, in this section we demonstrate that the tools 
of parameterized complexity analysis and, in particular, 
kernelization, may be beneficial in understanding and improving 
a known heuristic on the one side, and in providing 
a rigorous mathematical analysis on the other side. Here, we have 
examples in the context of graph completion problems, our key example here 
being the \textsc{Degree Anonymity} problem arising in the context of anonymizing social networks.

For many scientific disciplines, including the understanding of the spread of diseases in a globalized world or power consumption habits with impacts on energy efficiency, the availability of social network data becomes more and more important.
To respect privacy issues, there is a strong demand to anonymize the associated data in a preprocessing phase~\cite{FWCY10}.
If a graph contains only few vertices with some distinguished feature, then this might allow the identification (and violation of privacy) of the underlying real-world entities with that particular feature.
Hence, in order to ensure pretty good privacy and anonymity behavior, every vertex should share its feature with many other vertices. 
In a landmark paper, \citet{LT08} (also see \citet{CLT10} for an extended version) considered the vertex degrees as feature; see \citet{WYLC10} for other features considered in the literature. 
Correspondingly, a graph is called \emph{$\ell$-anonymous} if for each vertex there are at least~$\ell-1$ other vertices of the same degree.
Therein, different values of~$\ell$ reflect different privacy demands and the natural computational task arises to perform few changes to a graph in order to make it $\ell$-anonymous.

\newcommand{\inputToKDegAnonVariants}[1]{
		\tikzstyle{knoten}=[circle,draw,fill=black!20,minimum size=5pt,inner sep=2pt]
		\node[anchor=north west] at (0,\scale + 0.7) {\tworows{$\ell=\kValue$}{$k=#1$}};
		\node at (1,1) {};
		\node at (-1,-1) {};
		\node[knoten] (C) at (0,0) {};

		\foreach \i in {1,...,\n} {
			\node[knoten] (K-\i) at ({\scale * cos(360 * \i / \n - 3*(180 / \n))},{\scale * sin(360 * \i / \n - 3*(180 / \n))}) {};
			\path (C) edge[-] (K-\i);
		}
		\foreach \i in {2,...,\n}{
			\pgfmathtruncatemacro{\value}{\i - 1};
			\path (K-\i) edge[-] (K-\value);
		}
		\path (K-1) edge[-] (K-3);
}

\illustratedProbDefEnv{\defDecprob{\kDegAnon}
	{An undirected graph~$G=(V,E)$ and two integers~$k,\ell\in\N$.}
	{Is there an edge set~$S\subseteq \binom{V}{2}\setminus E$ of size at most~$k$ such that~$G+S$ is $\ell$-anonymous?} }
{
	\inputSolution{2.7}{-0.25}{-0.5}{3.5}{-0.5}{3.5}{
			\node[anchor=north west] at (2.2,2.6) {\tworows{$\ell=2$}{$k=1$}};

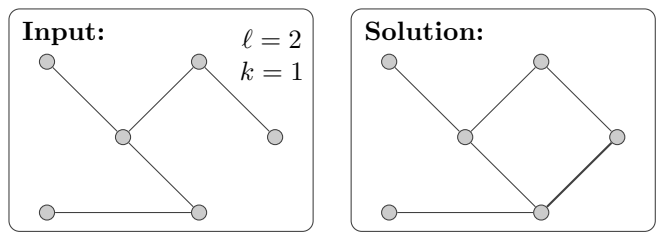
				\tikzstyle{knoten}=[circle,draw,fill=black!20,minimum size=5pt,inner sep=2pt]
		
				\node[knoten] (K-1) at (0,0) {};
				\node[knoten] (K-2) at (0,2) {};
				\node[knoten] (K-3) at (1,1) {};
				\node[knoten] (K-4) at (2,0) {};
				\node[knoten] (K-5) at (2,2) {};
				\node[knoten] (K-6) at (3,1) {};

				\foreach \i / \j in {1/4, 2/3, 3/4, 3/5, 5/6}{
					\path (K-\i) edge[-] (K-\j);
				}
	}{
				\tikzstyle{knoten}=[circle,draw,fill=black!20,minimum size=5pt,inner sep=2pt]
		
				\node[knoten] (K-1) at (0,0) {};
				\node[knoten] (K-2) at (0,2) {};
				\node[knoten] (K-3) at (1,1) {};
				\node[knoten] (K-4) at (2,0) {};
				\node[knoten] (K-5) at (2,2) {};
				\node[knoten] (K-6) at (3,1) {};

				\foreach \i / \j in {1/4, 2/3, 3/4, 3/5, 5/6}{
					\path (K-\i) edge[-] (K-\j);
				}
				\path (K-4) edge[-,thick] (K-6);
	}
}

The central parameterized complexity result for \textsc{Degree Anonymity} is that it has a polynomial-size problem kernel when parameterized by the maximum vertex degree~$\Delta$ of the input graph~\cite{HNNS15}.
In other words, there is a polynomial-time algorithm that transforms any input instance into an equivalent instance with $O(\Delta^7)$~vertices.
Indeed, one encounters a ``win-win'' situation when proving this result:
Liu and Terzi's heuristic strategy~\cite{LT08} finds an optimal solution when the size~$k$ of a minimum solution is larger than~$2\Delta^4$.
Hence, either one can solve the problem in polynomial time or the solution size is ``small''.
As a consequence, one can bound~$k$ in $O(\Delta^4)$ and, hence, a polynomial kernel for the combined parameter~$(\Delta,k)$ actually is also a polynomial kernel only for~$\Delta$.
While this kernelization directly implies fixed-parameter tractability for \textsc{Degree Anonymity} parameterized by~$\Delta$, 
there is also an  FPT~algorithm running in $O(\Delta^{O(\Delta^4)} + (k\ell+\Delta)\Delta k n)$ time.

The ideas behind the ``win-win'' situation generalize to further graph completion problems where the task is to insert edges so that the degree sequence of the resulting graph fulfills some prescribed property~$\Pi$~\cite{FNN16}.
Furthermore, an experimental evaluation of the usefulness of the theoretical results on the ``win-win'' situation delivered encouraging results even beyond the theoretical guarantees, that is, when~$k < 2\Delta^4$~\cite{HHN14,Nic15}.
This led to an enhancement of the heuristic due to \citet{LT08} which substantially improves on the previously known theoretical and empirical running times.
As for \textsc{Highly Connected Deletion}, previously known heuristic solutions could be substantially improved in terms of solution quality.

Finally, we mention in passing that making a graph $\ell$-anonymous
was studied from a parameterized point of view 
using also several other 
graph modification operations~\cite{BBHNW16,BFHNNT15,HT15}.
All these studies are of purely theoretical nature and there are 
only little positive
algorithmic results; links with 
heuristic algorithm design are missing. 

{F}rom a general perspective, the quest arising from the findings for
\textsc{Degree Anonymity} is to provide further examples where
parameterized complexity analysis sheds new light on known heuristics,
both theoretically and practically. A good starting point might be the
heuristic of \citet{LSB12} which clusters the vertices and then
anonymizes each cluster.  Here, the question is whether such a
practical link between anonymization and clustering could be
complemented with theoretical results. Obviously, these studies should
not be limited to problems arising in anonymization but to graph
modification problems from different application areas.

\section{Improving Heuristic Solutions with FPT~Algorithms}\label{sec:fast}
Local search is a generic algorithmic paradigm that yields good
heuristics for many optimization problems. The idea is to start with
any feasible solution and then search for a better one in the local
neighborhood of this solution. This search is continued until a
locally optimal solution is found. For graph modification problems, a
feasible solution~$S$ is any set of modification operations that
transforms the input graph into one that satisfies the graph property~$\Pi$. The local
neighborhood of~$S$ is usually defined as the sets of modification
operations that can be obtained by adding and removing at most~$k$
vertices from~$S$. This type of neighborhood is called~$k$-exchange
neighborhood.

An obvious approach to obtain more powerful local search algorithms is
to reduce the running time needed for searching the local
neighborhood. This could enable a local search algorithm to examine
larger neighborhoods and reduce the likelihood to remain in a locally
optimal but globally suboptimal solution. Usually, the size of the
$k$-exchange neighborhood in an $n$-vertex graph is upper-bounded by~$n^{f(k)}$ for some function~$f$. 
In parameterized algorithmics, a natural question is whether it is necessary to
consider all elements of this neighborhood or whether the neighborhood
can be searched faster, that is, in~$f(k)\cdot n^{O(1)}$ time.

For many vertex deletion problems this is not the
case~\cite{FFLRSV12}. For example, in the local search variant of
\textsc{Vertex Cover}, one is given a \emph{vertex cover}~$S$, that is, a
vertex set~$S$ such that deleting~$S$ from a graph~$G$ results in an
independent set. The task is to find a smaller vertex cover~$S'$ by adding and removing at most~$k$ vertices from~$S$.
\illustratedProbDefEnv{\defDecprob{\textsc{Local Search Vertex Cover}}
  {An undirected graph~$G=(V,E)$, a vertex cover~$S$ of~$G$, and an
    integer~$k\in\N$.}  {Is there a vertex cover~$S'\subseteq V$ such
    that~$|S'|<|S|$ and~$|(S\setminus S') \cup (S'\setminus S)|\le
    k$?} } { \inputSolution{2.9}{-0.25}{-0.5}{3.5}{-0.5}{3.8}{
    \node[anchor=north west] at (2.1,2.8) {\tworows{$|S|=5$}{$k=3$}};

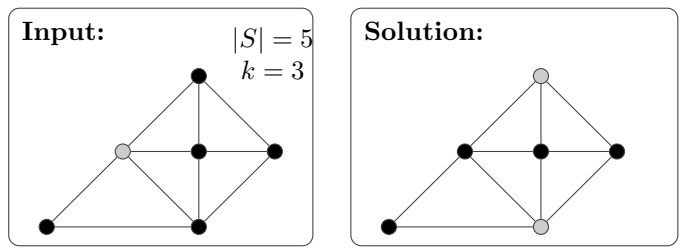
				\tikzstyle{knoten}=[circle,draw,fill=black!20,minimum size=5pt,inner sep=2pt]
				\tikzstyle{vcknoten}=[circle,draw,fill=black,minimum size=5pt,inner sep=2pt]
		
				\node[vcknoten] (K-1) at (0,0) {};
				\node[knoten] (K-2) at (1,1) {};
				\node[vcknoten] (K-3) at (2,1) {};
				\node[vcknoten] (K-4) at (2,0) {};
				\node[vcknoten] (K-5) at (2,2) {};
				\node[vcknoten] (K-6) at (3,1) {};

				\foreach \i / \j in {1/4, 1/2, 2/3, 2/4, 2/5, 3/4, 3/5, 3/6, 4/6, 5/6}{
					\path (K-\i) edge[-] (K-\j);
				}
	}{

				\tikzstyle{knoten}=[circle,draw,fill=black!20,minimum size=5pt,inner sep=2pt]
				\tikzstyle{vcknoten}=[circle,draw,fill=black,minimum size=5pt,inner sep=2pt]
		
				\node[vcknoten] (K-1) at (0,0) {};
				\node[vcknoten] (K-2) at (1,1) {};
				\node[vcknoten] (K-3) at (2,1) {};
				\node[knoten] (K-4) at (2,0) {};
				\node[knoten] (K-5) at (2,2) {};
				\node[vcknoten] (K-6) at (3,1) {};

				\foreach \i / \j in {1/4, 1/2, 2/3, 2/4, 2/5, 3/4, 3/5, 3/6, 4/6, 5/6}{
					\path (K-\i) edge[-] (K-\j);
				}	
             }
}
Unfortunately, unless W[1]${}={}$FPT, there is no FPT~algorithm for \textsc{Local Search Vertex Cover} parameterized by~$k$~\cite{FFLRSV12}. Positive results were obtained for special cases. For example, \textsc{Local Search Vertex Cover} and many other local search variants of vertex deletion problems are fixed-parameter tractable on planar graphs~\cite{FFLRSV12}. These
results, however, are based on the technique of locally bounded
treewidth. As a consequence, the resulting algorithms might not be
useful in practice.

Positive results were obtained for~\textsc{Feedback Arc Set in Tournaments} which is the problem of transforming a tournament, that is, a directed graph in which every pair of vertices is connected by exactly one of the two possible arcs, into an acyclic graph by a minimum number of arc deletions. 
Here, the local search problem is
fixed-parameter tractable. More precisely, given a set~$S$ of arc
deletions that makes a given tournament acyclic, it can be decided
in~$2^{O(\sqrt{k}\log k)}\cdot n^{O(1)}$ time whether there is a
set~$S'$ that can be obtained from~$S$ by adding and removing at
most~$k$ arcs~\cite{FLRS10}.

This positive result seems to be rooted in the
combinatorially restricted nature of tournaments and \emph{not} in the
fact that \textsc{Feedback Arc Set in Tournaments} is an arc
modification problem: The local search variant of the similarly simple
\textsc{Cluster Editing} problem is not fixed-parameter
tractable unless W[1]${}={}$FPT~\cite{DGKW14}.

Summarizing, the natural idea of parameterized local search faces two
major obstacles. The first obstacle is that, as discussed above, many
local search problems are probably not fixed-parameter tractable. The
second obstacle is that, so far, none of the parameterized local
search algorithms for graph modification problems have been shown to
be useful in practice. One encouraging result was obtained for
\textsc{Incremental List Coloring}~\cite{HN13}. Here, the input is a
graph with a list-coloring that colors all graph vertices except
one. The task is to obtain a list-coloring that also colors~$v$ and
disagrees with the old list-coloring on at most~$c$ vertices. Thus,
the new solution is searched within the neighborhood of the old
solution. This problem can be solved in~$k^c\cdot n^{O(1)}$ time
where~$k$ is the maximum size of any color list in the input. The
crucial observation is that this local search-like approach can be
embedded in a coloring heuristic that outperforms the standard greedy
coloring algorithm in terms of the coloring number. Since
\textsc{Incremental List Coloring} is W[1]-hard with respect to the
parameter~$c$, the key to success seems to be the consideration of the
combined parameter~$(k,c)$.

A goal for future research should thus be to obtain similar success
stories for local search variants of graph modification problems. As
demonstrated by \textsc{Incremental List Coloring}, one promising
route is the consideration of combined parameters. {F}rom a more
general perspective, the FPT algorithm for \textsc{Incremental List Coloring} and parameterized
local search have in common that they use the power provided by allowing
FPT running time---instead of polynomial running time---to improve
known heuristics. This approach, which has been coined
``turbo-charging heuristics''~\cite{DEF+14}, has close connections to
dynamic versions of hard graph problems~\cite{AEF+15,DEF+14}.

\section{Heuristic Tuning of Parameterized Algorithms}\label{sec:pbo}

Heuristics are often used to boost the performance of exact algorithms in practice.
A prominent example here is the branch-and-bound concept where heuristic lower and upper bounds restrict the search space for search-tree algorithms~\cite{MS08}.
Better heuristic bounds give a smaller search space and thus faster exact algorithms.
When analyzed in the classic complexity framework, the theoretical running time improvements due to the heuristic bounds are (if at all) marginal compared to the speed-ups observed in practice. 
Here, parameterized algorithmics can be used to give some theoretical explanation for experimental observations by using the above-guarantee parameterization~\cite{MR99}. 
As the name suggests, the parameter is the difference between the size of an optimal solution and a given lower bound. 
Fixed-parameter tractability with respect to the above-guarantee parameter then shows that the problem of finding a solution close to the lower bound is ``easy''. 
Thus, if the corresponding lower bound is close to the optimum, then the corresponding algorithm using this lower bound is fast---in practice \emph{and} in theory. 

An example for above-lower bound parameterization is \textsc{Vertex Cover}.
One lower bound on the size of a \textsc{Vertex Cover} is the value~$\ell$ of a linear programming (LP) relaxation.
The well-known LP relaxation is as follows:
\begin{align*} 
	\mbox{ \textbf{Minimize }} 	&&  \sum_{v\in V} x_v & && \\
	\mbox{\textbf{ subject to }} 	&& x_u+x_v  & \ge 1, && \forall \{u,v\}\in E \\
											&& x_v & \ge 0, && \forall v\in V. 
\end{align*}

It is known that in an optimal solution for the LP relaxation each variable has value 0, \nicefrac{1}{2}, or 1~\cite{NT74}.

\illustratedProbDefEnv{\defDecprob{\textsc{Vertex Cover Above LP}}
	{An undirected graph~$G=(V,E)$, an integer~$k \in\N$, and a rational number~$\ell \in \mathds{Q}$ denoting the value of the LP relaxation.}
	{Is there a vertex subset~$S \subseteq V$ of size at most~$k$ such that~$G[V \setminus S]$ is edgeless?} }
{
	\inputSolution{2.7}{-0.25}{-0.5}{3.5}{-0.5}{3.5}
	{
		\node[anchor=north west] at (1,2.6) {$k=3$, $\ell=2.5$};

		\tikzstyle{knoten}=[circle,draw,fill=black!20,minimum size=5pt,inner sep=2pt]
		
		\foreach[count=\i] \x / \y / \t / \yshift in {0/0.5/{\nicefrac{1}{2}}/0, 1/0/{\nicefrac{1}{2}}/-0.9, 1/1/{\nicefrac{1}{2}}/0, 2/0.5/{1}/0, 3/0/{0}/-0.9, 3/1/{0}/0} {
			\node[knoten,label={[xshift=0cm, yshift=\yshift cm]\t}] (K-\i) at (\x, \y +0.5) {};
		}

		\foreach \i / \j in {1/2, 1/3, 2/3, 3/4, 2/4, 4/5, 4/6}{
			\path (K-\i) edge[-] (K-\j);
		}
	}{
		\tikzstyle{knoten}=[circle,draw,minimum size=5pt,inner sep=2pt]
		
		\foreach[count=\i] \x / \y / \nodeStyle in {0/0.5/{black!20}, 1/0/{black}, 1/1/{black}, 2/0.5/{black}, 3/0/{black!20}, 3/1/{black!20}} {
			\node[knoten,fill=\nodeStyle] (K-\i) at (\x, \y +0.5) {};
		}

		\foreach \i / \j in {1/2, 1/3, 2/3, 3/4, 2/4, 4/5, 4/6}{
			\path (K-\i) edge[-] (K-\j);
		}
	}
}

\citet{LNRRS14} presented an algorithm solving \textsc{Vertex Cover Above LP} in~$2.32^{k-\ell} \cdot n^{O(1)}$ time.
On a high level, this algorithm starts with the lower bound and uses, after some preprocessing, a standard search-tree algorithm. 
Thus, a good lower bound allows not only in practice, but also in theory for an efficient algorithm solving \textsc{Vertex Cover}.
Moreover, the fixed-parameter tractability result now may help explaining why heuristics can successfully exploit the lower bound provided by the LP relaxation.

Another example for heuristic tuning of algorithms is programming by optimization~\cite{Hoo12}.
This is a helpful and powerful tool for developing fast implementations.
Here, the basic idea is that the implementation leaves open several 
design choices for different parts of the algorithm---these are 
settled later when training the algorithm with real-world instances.
Then, for the final configuration of the implementation, let a program choose from the alternatives in such a way that the performance is optimized on a representative set of instances. 
Here, the automated optimizer can give an answer to the following questions:
\begin{itemize}
	\item Given several alternative implementations for one subproblem (for example different sorting algorithms or different lower bounds), which one should be chosen?
	\item Should a certain data reduction rule be applied?  
	\item What are the ``best'' values for certain ``magic'' or ``hidden'' constants? For example, should a data reduction rule be applied in every second level of the search tree or every fourth level?
\end{itemize}
The programming by optimization approach has led to a state-of-the-art solver for \textsc{Cluster Editing}~\cite{HH15}. 
This solver combines one sophisticated data reduction rule and a branch-and-bound algorithm.
The solver outperforms previous algorithms which are based on integer linear programming (ILP) and pure branch-and-bound.
Thus, with the help of programming by optimization, implementations of parameterized algorithms may successfully compete with ILP-based algorithms.

On a high level, programming by optimization can be seen as a heuristic counterpart to parameterized algorithmics:
Parameterized algorithmics provides theoretical bounds on the running time of algorithms and the effectiveness of data reduction rules. These bounds depend on the parameter. Thus, to solve a problem for a specific type of data, one should measure different parameters and choose, based on this measurement, the most promising data reduction rules and algorithms.
With programming by optimization, this choice is made automatically, based on the performance of the algorithm on a given representative set of test instances.
Furthermore, the choice is not based on the values of parameters but directly on the efficiency of the corresponding algorithms on the test data. 

A goal for future research is to further increase the benefit obtained by combining the strengths of programming by optimization and parameterized algorithmics. This could be done, for example, by first providing several FPT algorithms for the same problem with different parameters and then using programming by optimization to find a good strategy to pick the best algorithm depending on the structure of an input instance. 
\section{Conclusions}\label{sec:concl}
As Karp~\cite{Kar11} pointed out,
one of the most pressing challenges
in theoretical computer science is to contribute to a better understanding
why many heuristics work so well in practice.
In particular, a formal footing of the construction
of heuristic algorithms is considered highly desirable.
This task is also closely connected to
(hidden) structure detection in real-world
input instances. We discussed several routes to a beneficial 
interaction between heuristic and parameterized algorithm design

%
To date, a clear majority of research results in parameterized
algorithmics is of purely theoretical nature. A natural way to
increase the practical impact of parameterized algorithmics is to seek
fruitful interactions with the field of heuristic algorithm
design.  We believe that particularly graph (modification) 
problems may be a forerunner in offering numerous fruitful
research opportunities in this direction.  

%



So far the strongest impact achieved by parameterized algorithmics on practical computing and heuristics is due to
kernelization, and polynomial-time 
data reduction techniques in general. Notably,
often data reduction rules seemingly not strong enough to provide 
kernelization results may still have strong practical impact.
Moreover, a general route for future research is to develop 
heuristic algorithms in parallel with performing a 
parameterized complexity analysis (particularly, in terms of kernelization). 
As results for graph modification 
problems in this direction demonstrate, there are good prospects to win
something in both worlds. 

Finally, in this paper we focused on NP-hard graph modification 
problems for illustrative examples. It goes without saying 
that our general remarks and observations are not limited 
to graph modification problems only but clearly extend to 
further graph problems and fields beyond, 
e.g.\ string algorithms~\cite{BHKN14} or 
computational social choice~\cite{BCFGNW14}.

\paragraph{Acknowledgment.}
We are grateful to Till Fluschnik and Vincent Froese for feedback to our manuscript.

\makeatletter
\renewcommand\bibsection%
{
  \section*{\refname
    \@mkboth{\MakeUppercase{\refname}}{\MakeUppercase{\refname}}}
}
\makeatother
{
	\bibliographystyle{abbrvnat}
	\bibliography{wg15-bibliography}
}


\end{document}